\documentclass[12pt]{article}

\usepackage[margin=1in]{geometry}
\usepackage{amsmath, amssymb, amsthm, authblk, hyperref, CJKutf8}

\newtheorem{lemma}{Lemma}
\newtheorem{theorem}{Theorem}

\theoremstyle{definition}
\newtheorem{example}{Example}

\usepackage[style=trad-unsrt, citestyle=numeric-comp, giveninits=true, doi=false, url=false, eprint=false, isbn=false]{biblatex}
\begin{document}

\title{Variance-based uncertainty relations}

\begin{CJK}{UTF8}{gbsn}

\author{Yichen Huang (黄溢辰)\thanks{yichuang@mit.edu} \thanks{Present address: Center for Theoretical Physics, Massachusetts Institute of Technology, Cambridge, Massachusetts 02139, USA.}}

\affil{Department of Physics, University of California, Berkeley, Berkeley, California 94720, USA}

\maketitle

\end{CJK}

\begin{abstract}

Uncertainty relations are fundamental in quantum mechanics. Here I propose state-independent variance-based uncertainty relations for two or more arbitrary observables in finite dimensional spaces. The uncertainty relations provide near-optimal lower bounds in some typical examples, and are useful for entanglement detection.

\end{abstract}

\section{Introduction}

Uncertainty relations are fundamental in quantum mechanics, underlying many conceptual differences between classical and quantum theories. They reveal by rigorous inequalities that incompatible observables cannot be measured to arbitrarily high precision simultaneously. They are useful in many areas related or even unrelated to quantum mechanics: entanglement detection \cite{HHHH09, GT09, HT03, Hua10, Hua10E}, quantum cryptography \cite{WW10}, signal processing \cite{CT06}, etc.

The Heisenberg uncertainty relation \cite{Rob29} reads
\begin{equation}
V(A)V(B)\ge\big|\langle\Psi|[A,B]|\Psi\rangle\big|^2/4,
\end{equation}
where $V(A),V(B)$ are the variances of the Hermitian operators $A,B$. The formulation of the Heisenberg uncertainty relation is not always satisfactory. For instance, the lower bound is trivial when $\langle\Psi|[A,B]|\Psi\rangle$ happens to be zero, which in finite dimensional spaces is always possible. Also, the lower bound depends on the state $|\Psi\rangle$, while in some applications a state-independent lower bound is needed \cite{HT03}. It is thus necessary to derive state-independent variance-based uncertainty relations. Note that no nontrivial state-independent lower bounds on the sum of the variances of two or more arbitrary observables are known, even if this formulation of uncertainty relations is precisely what is called for in literature \cite{HT03}.

Entropic uncertainty relations provide state-independent lower bounds on the sum of the entropies of two or more observables. They are especially useful in quantum cryptography \cite{WW10}. A lot of entropic uncertainty relations are proposed for observables in both finite \cite{WW10, MU88, dS08, BPPZ11, San95} and infinite \cite{Bec75, BM75, Hua11} dimensional spaces.

Here I propose state-independent variance-based uncertainty relations. Section \ref{sec:2} derives state-independent lower bounds on the sum of the variances of two or more arbitrary observables in finite dimensional spaces. Section \ref{sec:3} concludes with some typical examples in which the lower bounds are near-optimal and with the application to entanglement detection. The Appendix briefly surveys entropic uncertainty relations, which are useful in formulating the main theorems.

\section{Main theorems} \label{sec:2}

Let $\{\{|a_i^j\rangle|j=1,2,\ldots,n\}|i=1,2,\ldots,m\}$ be a set of $m$ orthonormal bases of an $n$-dimensional Hilbert space, and $A_i=\sum_{j=1}^na_i^j|a_i^j\rangle\langle a_i^j|$ be $m$ Hermitian operators, where $a_i^j\in\mathbb R$ and $|a_i^j\rangle$ are the eigenvalues and the eigenvectors of $A_i$, respectively. For notational simplicity, let $p_i^j=|\langle a_i^j|\Psi\rangle|^2$. The Shannon entropy of $A_i$ is 
\begin{equation}
H(A_i)=-\sum_{j=1}^np_i^j\ln p_i^j.
\end{equation}
From now on, assume the entropic uncertainty relation
\begin{equation}
\sum_{i=1}^mH(A_i)\ge C
\end{equation}
Let $\alpha$ be a variational parameter. All the inequalities below hold for any $\alpha\in\mathbb R$.

\begin{lemma}
\begin{equation}
\alpha V(A_i)\ge H(A_i)-\ln\sum_{j=1}^ne^{-\alpha(a_i^j-\langle A_i\rangle)^2}.
\label{eq:lemma}
\end{equation}
\end{lemma}

\begin{proof}
Using the basic inequality $e^x\ge1+x$ with $x=-\alpha(a_i^k-\langle A_i\rangle)^2-\ln(p_i^k\sum_{j=1}^ne^{-\alpha(a_i^j-\langle A_i\rangle)^2})$,
\begin{align}
1&=\sum_{k=1}^np_i^k\times\frac{e^{-\alpha(a_i^k-\langle A_i\rangle)^2}}{p_i^k\sum_{j=1}^{n}e^{-\alpha(a_i^j-\langle A_i\rangle)^2}}\ge\sum_{k=1}^np_i^k\left(1-\alpha(a_i^k-\langle A_i\rangle)^2-\ln\left(p_i^k\sum_{j=1}^{n}e^{-\alpha(a_i^j-\langle A_i\rangle)^2}\right)\right)\nonumber\\
&=\sum_{k=1}^n p_i^k-\alpha\sum_{k=1}^n p_i^k(a_i^k-\langle A_i\rangle)^2-\sum_{k=1}^n p_i^k\ln p_i^k-\left(\sum_{k=1}^np_i^k\right)\ln \sum_{j=1}^{n}e^{-\alpha(a_i^j-\langle A_i\rangle)^2}\nonumber\\
&=1-\alpha V(A_i)+H(A_i)-\ln\sum_{j=1}^{n}e^{-\alpha(a_i^j-\langle A_i\rangle)^2}.
\end{align}
\end{proof}

Add (\ref{eq:lemma}) for $i=1,2,\ldots,m$:
\begin{theorem} [state-dependent variance-based uncertainty relation] \label{thm1}
\begin{equation}
\alpha\sum_{i=1}^mV(A_i)\ge C-\sum_{i=1}^m\ln\sum_{j=1}^ne^{-\alpha(a_i^j-\langle A_i\rangle)^2}.
\end{equation}
\end{theorem}

Note $\min_ka_i^k\le\langle A_i\rangle\le\max_ka_i^k$:
\begin{theorem} [state-independent variance-based uncertainty relation] \label{thm2}
\begin{equation} \label{eq:multi}
\alpha\sum_{i=1}^mV(A_i)\ge C-\sum_{i=1}^m\ln\max_{\min_ka_i^k\le\beta_i\le\max_ka_i^k}\sum_{j=1}^ne^{-\alpha(a_i^j-\beta_i)^2}.
\end{equation}
\end{theorem}

Usually the maximum in (\ref{eq:multi}) cannot be computed analytically, but it is easy to compute numerically. Theorems \ref{thm1} and \ref{thm2} hold more generally for positive-operator valued measures provided with the corresponding entropic uncertainty relations for positive-operator valued measures.

\section{Examples, applications, and conclusions} \label{sec:3}

A set of $m$ orthonormal bases of an $n$-dimensional Hilbert space is mutually unbiased if
\begin{equation}
\big|\langle a_i^j|a_l^k\rangle\big|=\frac{1-\delta_{il}}{\sqrt n}+\delta_{il}\delta_{jk}
\end{equation}
for $i,l=1,2,\ldots,m;j,k=1,2,\ldots,n$. Mutually unbiased bases are especially useful in quantum cryptography \cite{WW10}.

\begin{example}
The eigenvectors of the Pauli matrices are mutually unbiased bases. (\ref{eq:e2}) shows $C=2\ln 2$. The optimal choice of the variational parameter is $\alpha=0.597$:
\begin{equation} \label{eq:b1}
V(\sigma_x)+V(\sigma_y)+V(\sigma_z)>1.7243.
\end{equation}
\end{example}

\begin{example}
The eigenvectors of the matrices 
\begin{gather}
\sigma_0=\begin{pmatrix}1&0&0\\0&-1&0\\0&0&0\end{pmatrix},\quad\sigma_1=\frac{i}{\sqrt3}\begin{pmatrix}0&-1&1\\1&0&-1\\-1&1&0\end{pmatrix},\nonumber\\
\sigma_2=\frac{1}{\sqrt3}\begin{pmatrix}0&\omega^{11}&\omega^9\\\omega&0&\omega^7\\\omega^3&\omega^5&0\end{pmatrix},\quad\sigma_3=\frac{1}{\sqrt3}\begin{pmatrix}0&\omega&\omega^3\\\omega^{11}&0&\omega^5\\\omega^9&\omega^7&0\end{pmatrix}
\end{gather}
are mutually unbiased bases, where $\omega=e^{\pi i/6}$. (\ref{eq:e2}) shows $C=4\ln2$. The optimal choice of the variational parameter is $\alpha=1.92$:
\begin{equation} \label{eq:b2}
V(\sigma_0)+V(\sigma_1)+V(\sigma_2)+V(\sigma_3)>0.9083.
\end{equation}
\end{example}

The lower bounds in (\ref{eq:b1}) and (\ref{eq:b2}) are near-optimal since the optimal lower bounds are $2$ and $1$, respectively.

The state-independent variance-based uncertainty relation (\ref{eq:multi}) is precisely in the formulation that is called for in the literature of entanglement detection \cite{HT03}. Entanglement detection is an important problem as entanglement plays a key role in quantum information. Among many detection schemes \cite{HHHH09, GT09}, the entanglement criterion based on local uncertainty relations \cite{HT03} is well known. It is a necessary condition for separability:
\begin{equation}
\sum_i V(A_i+B_i)\ge U_A+U_B,
\end{equation}
where $A_i$ and $B_i$ are Hermitian operators on the first and the second subspaces, respectively, with $U_A,U_B$ given by the state-independent variance-based uncertainty relations:
\begin{equation}
\sum_i V(A_i)\ge U_A,\quad\sum_i V(B_i)\ge U_B.
\end{equation}
Section II in Ref. \cite{HT03} appreciates the importance of $U_A,U_B$ in formulating the criterion, but how to evaluate $U_A,U_B$ remains unclear, which is very inconvenient for users. The state-independent variance-based uncertainty relation (\ref{eq:multi}) precisely fills the gap.

In conclusion, I have proposed state-independent variance-based uncertainty relations for two or more arbitrary observables in finite dimensional spaces. The uncertainty relations provide near-optimal lower bounds in some typical examples, and are useful for entanglement detection.

\appendix

\section{Brief survey of entropic uncertainty relations}

Entropic uncertainty relations are useful in formulating the main theorems. See Ref. \cite{WW10} for a recent comprehensive survey.

The entropic uncertainty relation for two Hermitian operators \cite{MU88} reads
\begin{equation} \label{eq:mu}
H(A_1)+H(A_2)\ge-2\ln\max_{1\le j,k\le n}\big|\langle a_1^j|a_2^k\rangle\big|,
\end{equation}
and the lower bound is improved for $\max_{1\le j,k\le n}|\langle a_1^j|a_2^k\rangle|\ge1/\sqrt2$ \cite{dS08, BPPZ11}. (\ref{eq:mu}) reduces to
\begin{equation}
H(A_1)+H(A_2)\ge\ln n
\end{equation}
for mutually unbiased bases. The entropic uncertainty relation for $n+1$ mutually unbiased bases \cite{San95} reads
\begin{equation} \label{eq:e2}
\sum_{i=1}^{n+1}H(A_i)\ge\left[\frac{n+1}{2}\right]\ln\left[\frac{n+1}{2}\right]+\left[\frac{n+2}{2}\right]\ln\left[\frac{n+2}{2}\right],
\end{equation}
where $[\cdots]$ denotes the floor function. The entropic uncertainty relation for the position and the momentum operators $x,p$ \cite{Bec75, BM75} reads
\begin{equation}
H(x)+H(p)\ge1+\ln\pi.
\end{equation}
Its multimode generalization is given in Ref. \cite{Hua11}.

\printbibliography

\end{document}